\documentclass[seceq]{ptptex}

\usepackage{graphicx}

\def\bG{{\bf G}}

\def\bk{{\bf k}}

\def\bp{{\bf p}}

\def\bG{{\bf G}}
\def\bL{{\bf L}}

\def\bR{{\bf R}}
\def\bS{{\bf S}}

\def\bU{{\bf U}}

\def\b0{{\bf 0}}

\def\cO{{\cal O}}

\def\up{\uparrow}
\def\down{\downarrow}

\def\lra{\leftrightarrow}

\def\eps{\epsilon}

\def\Gam{\Gamma}

\def\Lam{\Lambda}

\def\sg{\sigma}
\def\Sg{\Sigma}
\def\bGam{{\bf\Gamma}}
\def\bSg{{\bf\Sigma}}

\def\phib{\bar\phi}
\def\psib{\bar\psi}

\title{Parametrization of Nambu vertex in a singlet superconductor}

\author{Andreas \textsc{Eberlein}$^1$ and Walter \textsc{Metzner}$^1$}

\inst{
$^1$Max-Planck-Institute for Solid State Research, D-70569 Stuttgart, 
Germany}

\abst{We analyze general properties of the effective Nambu two-particle 
 vertex and its renormalization group flow in a spin-singlet 
 superconductor. In a fully spin-rotation invariant form the Nambu 
 vertex can be expressed by only three distinct components.
 Solving exactly the flow of a mean-field model with reduced BCS and
 forward scattering interactions, we gain insight into the singularities
 in the momentum and energy dependences of the vertex at and below the
 critical energy scale for superconductivity.
 Using a decomposition of the vertex in various interaction channels, we
 manage to isolate singular momentum and energy dependences in only
 one momentum and energy variable for each term, such that the 
 singularities can be efficiently parametrized.}

\begin{document}

\maketitle

\section{Introduction}

In the last decade the functional renormalization group (fRG) 
has been established as a valuable source of new approximation
schemes for interacting Fermi systems \cite{Metzner05}.
These approximations are obtained as truncations of an exact 
functional flow equation which yields the flow from the bare 
microscopic action to the quantum effective action as a function 
of a decreasing energy scale $\Lambda$.\cite{Berges02}

Most interacting Fermi systems undergo a spontaneous symmetry
breaking at sufficiently low temperatures (sometimes only at
$T=0$). In the functional flow equation, the common types
of spontaneous symmetry breaking such as superconductivity or
magnetic order are associated with a divergence of the effective
two-particle interaction at a finite scale $\Lambda_c$, in a 
certain momentum channel.\cite{Zanchi00,Halboth00,Honerkamp01} 
To continue the flow below the scale $\Lambda_c$, an appropriate
order parameter has to be introduced.

One possibility is to decouple the interaction by a bosonic order
parameter field, via a Hubbard-Stratonovich transformation, and
to study the coupled flow of the fermionic and order parameter
fields.
In this way order parameter fluctuations and also their 
interactions can be conveniently treated. 
This route to symmetry breaking in the fRG framework has been 
explored already in several works on antiferromagnetism \cite{Baier04} 
and superconductivity.\cite{Birse05,Diehl07,Strack08}
Typically the bare microscopic interaction can be decoupled in
various channels, but the results obtained from truncated flow
equations depend on the choice of the Hubbard-Stratonovich field.
This ambiguity is particularly serious in the case of competing
instabilities with distinct order parameters.

It is therefore worthwhile to explore also purely fermionic
flows in the symmetry-broken phase. This can be
done by adding an infinitesimal symmetry breaking term to the
bare action, which is promoted to a finite order parameter below
the scale $\Lambda_c$.\cite{Salmhofer04}
A relatively simple one-loop truncation of the exact fRG flow
equation was shown to yield an {\em exact} description of symmetry
breaking for mean-field models such as the reduced BCS model,
although the effective two-particle interactions diverge at
the critical scale $\Lambda_c$.\cite{Salmhofer04,Gersch05}
It turned out that the same truncation, with a very simple
parametrization of the effective two-particle vertex,
provides also surprisingly accurate results for the superconducting 
gap in the ground state of the two-dimensional attractive Hubbard 
model at weak coupling.\cite{Gersch08}

The purpose of the present work is to further develop the
fermionic fRG route to spontaneous symmetry breaking, focusing
on the case of singlet superconductivity as a prototype for a 
broken continuous symmetry.
We stay with the one-loop truncation used already in the previous
works, but we improve on the parametrization of the effective
two-particle interaction in two directions.
First, we make full use of spin-rotation invariance to reduce
the number of independent Nambu components of the two-particle 
vertex to a minimum.
Second, we pave the way for an efficient parametrization of
singular momentum and energy dependences by extending a recently
proposed channel decomposition \cite{Husemann09} of the two-particle 
vertex to the symmetry-broken state of a superconductor.
We gain insight into the singularity structure of the vertex by 
analysing the exact fRG flow of an extended mean-field model 
featuring not only reduced BCS interactions, but also charge and 
spin forward scattering.

The paper is structured as follows. 
In {\S}2 we review the one-loop truncation of the exact flow 
equation, and we introduce the Nambu representation for the 
two-particle vertex in a spin-singlet superconductor.
Available symmetries are exploited in {\S}3, with the aim of
parametrizing the Nambu two-particle vertex by a minimal 
number of independent components. 
In {\S}4 we compute the exact renormalization group flow for
the reduced model with BCS and forward scattering interactions,
paying particular attention to the singularities of the Nambu 
vertex developed in the course of the flow.
The channel decomposition introduced by Husemann and Salmhofer
\cite{Husemann09} for an efficient parametrization of momentum
and frequency dependences in the normal state is extended to
singlet superconductors in {\S}5.
We finally conclude with a summary of our results in {\S}6.

\section{Truncated flow equations}

We consider a system of interacting spin-$\frac{1}{2}$ fermions
with a single-particle basis given by states with a momentum
$\bk$, a spin orientation $\sg \in \{\up,\down\}$, and a kinetic
energy $\eps_{\bk}$.
The system is specified by an action
\begin{equation}
 S[\psi,\psib] = 
 \sum_{k,\sg} (-ik_0 + \xi_{\bk}) \, \psib_{k\sg} \psi_{k\sg} +
 V[\psi,\psib] \; ,
\end{equation}
where $k = (k_0,\bk)$ contains the Matsubara frequency $k_0$ in
addition to the momentum; 
$\psib_{k\sg}$ and $\psi_{k\sg}$ are Grassmann variables 
associated with creation and annihilation operators, respectively;
$\xi_{\bk} = \eps_{\bk} - \mu$ is the single-particle energy
relative to the chemical potential, and $V[\psi,\psib]$ is a 
two-particle interaction of the form
\begin{equation}
 V[\psi,\psib] = \frac{1}{4}
 \sum_{k_1,\dots,k_4} \sum_{\sg_1,\dots,\sg_4}
 V_{\sg_1\sg_2\sg_3\sg_4}(k_1,k_2,k_3,k_4) \,
 \psib_{k_1\sg_1}  \psib_{k_2\sg_2} \psi_{k_3\sg_3} \psi_{k_4\sg_4}
 \; .
\end{equation}
Here and in the following the usual temperature and volume factors
are incorporated in the summation symbols.

The starting point for our analysis of the interacting Fermi system is 
an exact flow equation \cite{Wetterich93,Salmhofer01} for the effective 
action $\Gam^{\Lam}[\psi,\psib]$, that is, the generating functional
for one-particle irreducible vertex functions in the presence of an
infrared cutoff $\Lam$. The cutoff is implemented by endowing the bare
propagator with a regulator function. $\Gam^{\Lam}[\psi,\psib]$ 
interpolates smoothly between the bare action 
$\Gam^{\Lam_0}[\psi,\psib] = S[\psi,\psib]$ and the final effective 
action $\Gam[\psi,\psib]$ in the limit $\Lam \to 0$.
Spontaneous symmetry breaking can be treated by adding an infinitesimal
symmetry breaking field to the bare action, which is then promoted to
a finite order parameter in the course of the flow.\cite{Salmhofer04}

Expanding the exact functional flow equation for $\Gam^{\Lam}[\psi,\psib]$
in powers of $\psi$ and $\psib$, one obtains a hierarchy of flow
equations for the n-particle vertex functions $\Gam^{(2n)\Lam}$.
We truncate the hierarchy at the two-particle level, including however
self-energy corrections due to contractions of three-particle terms.
\cite{Katanin04}
This truncation, which is sketched diagrammatically in Figs.~1 and 2,
\begin{figure}
\centerline{\includegraphics[width=3.5cm]{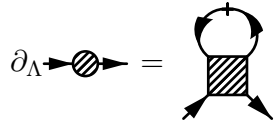}}
\caption{Diagrammatic representation of the flow equation for the
 self-energy. The slashed line represents the single-scale propagator
 $S^{\Lam}$.}
\end{figure}
\begin{figure}
\centerline{\includegraphics[width=9cm]{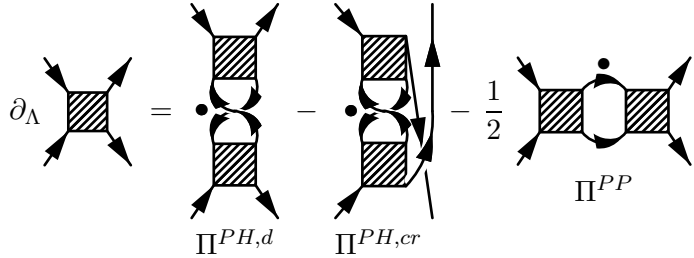}}
\caption{Diagrammatic representation of the flow equation for the
 two-particle vertex. The dots near the internal lines indicate
 differentiation of the propagator products with respect to the 
 scale $\Lam$.}
\end{figure}
was used in all previous studies of symmetry breaking from a fermionic
flow.\cite{Salmhofer04,Gersch05,Gersch08}
The flow of the self-energy $\Sg^{\Lam}$ is determined by a tadpole
contraction of the two-particle vertex $\Gam^{(4)\Lam}$ and the
socalled single-scale propagator 
\begin{equation}
 S^{\Lam} = 
 \partial_{\Lam} G^{\Lam} |_{\Sg^{\Lam} \; {\rm fixed}} \; ,
\end{equation}
where $G^{\Lam}$ is the infrared regularized propagator.
The latter is defined by
$(G^{\Lam})^{-1} = \Gam^{(2)\Lam} + R^{\Lam}$ with a suitable
regulator function $R^{\Lam}$.
The flow of the two-particle vertex $\Gam^{(4)\Lam}$ is determined
by a one-loop diagram involving $\Gam^{(4)\Lam}$ itself and the
total cutoff derivative of the regularized propagator 
\begin{equation}
 \partial_{\Lam} G^{\Lam} =
 S^{\Lam} - G^{\Lam} (\partial_{\Lam} \Sg^{\Lam}) G^{\Lam} \; .
\end{equation}
The truncated system of flow equations described in Figs.~1 and 2 has
the merit of solving mean-field models such as the reduced BCS model
\cite{Salmhofer04} and the reduced charged density wave model 
\cite{Gersch05} exactly.

We now specify to singlet superconductivity, where the continuous
$U(1)$ symmetry associated with charge conservation is spontaneously
broken, while spin-rotation invariance remains conserved.
It is useful to use Nambu spinors $\phi_{ks}$ and $\phib_{ks}$
instead of $\psi_{k\sg}$ and $\psib_{k\sg}$ as a basis. 
The two basis sets are related by
\begin{equation}
 \phib_{k+} = \psib_{k\up}, \quad 
 \phi_{k+}  = \psi_{k\up}, \quad
 \phib_{k-} = \psi_{-k\down}, \quad 
 \phi_{k-}  = \psib_{-k\down} \; .
\label{eq:phipsi}
\end{equation}
The effective action as a functional of the Nambu fields,
truncated beyond two-particle terms, has the form
\begin{eqnarray}
 \Gam^{\Lam}[\phi,\phib] &=& 
 \Gam^{(0)\Lam} + 
 \sum_k \sum_{s_1,s_2} \Gam_{s_1s_2}^{(2)\Lam}(k) \,
 \phib_{ks_1} \phi_{ks_2} \nonumber \\
 &+& \frac{1}{4} \sum_{k_1,\dots,k_4} \sum_{s_1,\dots,s_4}
 \Gam_{s_1s_2s_3s_4}^{(4)\Lam}(k_1,k_2,k_3,k_4) \,
 \phib_{k_1s_1} \phib_{k_2s_2} \phi_{k_3s_3} \phi_{k_4s_4} \; ,
\end{eqnarray}
where $\Gam^{(0)\Lam}$ yields the grandcanonical potential.
Due to spin-rotation invariance only terms with an equal
number of $\phi$ and $\phib$ fields contribute.
The regularized Nambu propagator $\bG^{\Lam}$ is related
to $\bGam^{(2)\Lam}$ by
$(\bG^{\Lam})^{-1} = \bGam^{(2)\Lam} + \bR^{\Lam}$, and 
can be written as a $2 \times 2$ matrix of the form
\begin{equation}
 \bG^{\Lam}(k) = 
 \left( \begin{array}{cc}
 G_{++}^{\Lam}(k) & G_{+-}^{\Lam}(k) \\
 G_{-+}^{\Lam}(k) & G_{--}^{\Lam}(k)
 \end{array} \right) =
 \left( \begin{array}{cc}
 G^{\Lam}(k) & F^{\Lam}(k) \\
 F^{*\Lam}(k) & -G^{\Lam}(-k)
 \end{array} \right) \; .
\end{equation}
Note that the anomalous propagator $F^{\Lam}(k)$ is a 
symmetric function of $k_0$ and $\bk$, due to spin-rotation
and reflection invariance.

For a singlet superconductor, the flow equation for the
Nambu self-energy $\bSg$, shown graphically in Fig.~1, is 
given by
\begin{equation} \label{floweq_selfen}
 \frac{d}{d\Lam} \Sg_{s_1s_2}^{\Lam}(k) =
 - \sum_k \sum_{s_3,s_4} S_{s_4s_3}^{\Lam}(k')
 \Gam^{(4)\Lam}_{s_1s_3s_4s_2}(k,k',k',k) \; ,
\end{equation}
and the flow equation for the Nambu vertex (Fig.~2) reads
\begin{eqnarray} \label{floweq_vertex}
 \frac{d}{d\Lam}
 \Gam^{(4)\Lam}_{s_1s_2s_3s_4}(k_1,k_2,k_3,k_4) 
 &=& 
 \Pi^{PH,d}_{s_1s_2s_3s_4}(k_1,k_2,k_3,k_4) - 
 \Pi^{PH,cr}_{s_1s_2s_3s_4}(k_1,k_2,k_3,k_4) \nonumber \\ 
 &-& 
 \frac{1}{2} \Pi^{PP}_{s_1s_2s_3s_4}(k_1,k_2,k_3,k_4) \; ,
\end{eqnarray}
where
\begin{eqnarray} \label{bubbles}
 \Pi^{PH,d}_{s_1s_2s_3s_4}(k_1,k_2,k_3,k_4) &=&
 \sum_{p,q} \sum_{s'_1,\dots,s'_4}
 \frac{d}{d\Lam}
 [G^{\Lam}_{s'_1s'_2}(p) G^{\Lam}_{s'_3s'_4}(q)] \nonumber \\
 &\times& 
 \Gam^{(4)\Lam}_{s_1s'_2s'_3s_4}(k_1,p,q,k_4)
 \Gam^{(4)\Lam}_{s'_4s_2s_3s'_1}(q,k_2,k_3,p) \; , \\
 \Pi^{PH,cr}_{s_1s_2s_3s_4}(k_1,k_2,k_3,k_4) &=&
 \sum_{p,q} \sum_{s'_1,\dots,s'_4}
 \frac{d}{d\Lam}
 [G^{\Lam}_{s'_1s'_2}(p) G^{\Lam}_{s'_3s'_4}(q)] \nonumber \\
 &\times& 
 \Gam^{(4)\Lam}_{s_2s'_2s'_3s_4}(k_2,p,q,k_4)
 \Gam^{(4)\Lam}_{s'_4s_1s_3s'_1}(q,k_1,k_3,p) \; , \\
 \Pi^{PP}_{s_1s_2s_3s_4}(k_1,k_2,k_3,k_4) &=&
 \sum_{p,q} \sum_{s'_1,\dots,s'_4}
 \frac{d}{d\Lam}
 [G^{\Lam}_{s'_1s'_2}(p) G^{\Lam}_{s'_3s'_4}(q)] \nonumber \\
 &\times& 
 \Gam^{(4)\Lam}_{s_1s_2s'_3s'_1}(k_1,k_2,q,p)
 \Gam^{(4)\Lam}_{s'_2s'_4s_3s_4}(p,q,k_3,k_4) \; .
\end{eqnarray}
Note that the particle-particle terms in the Nambu 
representation contain particle-hole contributions in the 
original fermion basis and vice versa. In particular, the
original particle-particle contribution driving the 
superconducting instability is contained in the Nambu
particle-hole diagrams.

We assume translation invariance such that the one-particle
propagator and self-energy depend only on a single energy
and momentum variable $k$, while the two-particle quantities
are non-zero only if $k_1 + k_2 = k_3 + k_4$, and can
therefore be parametrized by three independent energy
and momentum variables.

\section{Constraints from symmetries}

In this section we exploit the available symmetries to reduce
the number of independent components of the two-particle 
vertex in the superconducting phase to a minimum.
In any case, the effective action is invariant under translations, 
spin rotations and spatial inversions.
In most singlet superconductors, also time reversal invariance
remains unbroken.

In addition to the normal interaction involving two creation 
and two annihilation operators ($\psib\psib\psi\psi$)
there are also {\em anomalous vertices} corresponding to 
operator products $\psib\psib\psib\psib$ + conjugate and 
$\psib\psib\psib\psi$ + conjugate.\cite{Salmhofer04,Gersch08}
We now write down manifestly spin-rotation invariant forms
for each of these terms in the $\psi$-basis, and then 
translate to the Nambu representation.

A spin-rotation invariant normal interaction can be written
as \cite{Salmhofer01}
\begin{eqnarray} \label{gamma22}
 \Gam^{(2+2)}[\psi,\psib] &=& 
 \frac{1}{4} \sum_{k_i,\sg_i} \left[ 
 V(k_1,k_2,k_3,k_4) \delta_{\sg_1\sg_4} \delta_{\sg_2\sg_3} -
 V(k_1,k_2,k_4,k_3) \delta_{\sg_1\sg_3} \delta_{\sg_2\sg_4}
 \right] \nonumber \\
 &\times&
 \psib_{k_1\sg_1} \psib_{k_2\sg_2} \psi_{k_3\sg_3} \psi_{k_4\sg_4}
 \; .
\end{eqnarray}
Here and in the remainder of this section we suppress the
superscript $\Lam$ denoting the scale dependence.
Alternatively one may write $\Gam^{(2+2)}[\psi,\psib]$ as a
sum of a spin singlet and a spin triplet component \cite{Halboth00}
\begin{eqnarray}
 \Gam^{(2+2)}[\psi,\psib] &=& 
 \frac{1}{4} \sum_{k_i,\sg_i} \left[ 
 V^S(k_1,k_2,k_3,k_4) S_{\sg_1\sg_2\sg_3\sg_4} +
 V^T(k_1,k_2,k_3,k_4) T_{\sg_1\sg_2\sg_3\sg_4}
 \right] \nonumber \\
 &\times&
 \psib_{k_1\sg_1} \psib_{k_2\sg_2} \psi_{k_3\sg_3} \psi_{k_4\sg_4}
 \; ,
\end{eqnarray}
where 
$S_{\sg_1\sg_2\sg_3\sg_4} = 
 \frac{1}{2} (\delta_{\sg_1\sg_4} \delta_{\sg_2\sg_3}
 - \delta_{\sg_1\sg_3} \delta_{\sg_2\sg_4})$,
$T_{\sg_1\sg_2\sg_3\sg_4} = 
 \frac{1}{2} (\delta_{\sg_1\sg_4} \delta_{\sg_2\sg_3}
 + \delta_{\sg_1\sg_3} \delta_{\sg_2\sg_4})$, and
\begin{eqnarray}
 V^S(k_1,k_2,k_3,k_4) &=& 
 V(k_1,k_2,k_3,k_4) + V(k_1,k_2,k_4,k_3) \; , \\
 V^T(k_1,k_2,k_3,k_4) &=& 
 V(k_1,k_2,k_3,k_4) - V(k_1,k_2,k_4,k_3) \; .
\end{eqnarray}
Time reversal invariance \cite{Banyai94} and charge conjugation
(corresponding to hermiticity of the underlying Hamiltionian)
yield the following relations for $V(k_1,k_2,k_3,k_4)$:
\begin{eqnarray}
 V(k_1,k_2,k_3,k_4) &=& 
 V(k_4^{\cal T},k_3^{\cal T},k_2^{\cal T},k_1^{\cal T})  \quad
 \mbox{(time reversal)} \; , \\
 V(k_1,k_2,k_3,k_4) &=& 
 V^*(k_4^*,k_3^*,k_2^*,k_1^*)  \quad \; \,
 \mbox{(charge conjugation)} \; ,
\end{eqnarray}
where $k^{\cal T} = (k_0,-\bk)$ and $k^* = (-k_0,\bk)$,
and the same for $V^S$ and $V^T$.
Note that spatial inversions also transform $k$ to $k^{\cal T}$,
but without permuting the momenta in $V(k_1,k_2,k_3,k_4)$.
The invariance of the action under the exchange of identical 
particles yields separate symmetry relations for $V^S$ and $V^T$
under $k_1 \lra k_2$ and $k_3 \lra k_4$:
\begin{eqnarray}
 V^S(k_1,k_2,k_3,k_4) =& \phantom- V^S(k_2,k_1,k_3,k_4) \, &=  
 \phantom- V^S(k_1,k_2,k_4,k_3) \; , \\
 V^T(k_1,k_2,k_3,k_4) =& - V^T(k_2,k_1,k_3,k_4) \, &= 
 - V^T(k_1,k_2,k_4,k_3) \; ,
\end{eqnarray}
while $V$ obeys only
\begin{equation}
 V(k_1,k_2,k_3,k_4) = V(k_2,k_1,k_4,k_3) \; .
\end{equation}

A spin-rotation invariant ansatz for the anomalous interactions 
can be constructed systematically by requiring that commutators 
with the three components of the total spin operator vanish.

A spin-rotation invariant anomalous interaction involving four
creation or four annihilation operators can be written in the 
form
\begin{eqnarray} \label{gamma40}
 \Gam^{(4+0)}[\psi,\psib] =&& \nonumber \\
 \frac{1}{8} \sum_{k_i} &&
 \Big\{ W^{S}(k_1,k_2,k_3,k_4) 
 (\psib_{k_1\up}\psib_{k_2\down} - \psib_{k_1\down}\psib_{k_2\up})
 (\psib_{k_3\up}\psib_{k_4\down} - \psib_{k_3\down}\psib_{k_4\up})
 \nonumber \\
  - && \,
 W^{T}(k_1,k_2,k_3,k_4) \big[
 (\psib_{k_1\up}\psib_{k_2\down} + \psib_{k_1\down}\psib_{k_2\up})
 (\psib_{k_3\up}\psib_{k_4\down} + \psib_{k_3\down}\psib_{k_4\up})
 \nonumber \\[2mm]
 && -2 
 (\psib_{k_1\up} \psib_{k_2\up} \psib_{k_3\down} \psib_{k_4\down} +
  \psib_{k_1\down} \psib_{k_2\down} \psib_{k_3\up} \psib_{k_4\up})
 \big] + {\rm conj.} \Big\} \; ,
\end{eqnarray}
consisting of a singlet ($S$) and a triplet ($T$) term, which are
separately invariant under spin rotations.
Charge conjugated terms are denoted by ''conj.''.
Time reversal invariance yields a relation between $W^{S,T}$ 
and $W^{S,T*}$, which assumes the simple form
\begin{equation}
 W^{S,T*}(k_1,k_2,k_3,k_4) = 
 W^{S,T}(-k_1,-k_2,-k_3,-k_4)  \; ,
\end{equation}
if the order parameter (gap function) is chosen real.
Furthermore, the functions $W^{S}(k_1,k_2,k_3,k_4)$ and
$W^{T}(k_1,k_2,k_3,k_4)$ are symmetric and antisymmetric 
under the exchange $k_1 \lra k_2$ or $k_3 \lra k_4$, respectively, 
and symmetric under the simultaneous exchange  $k_1 \lra k_3$ and 
$k_2 \lra k_4$:
\begin{eqnarray}
 W^S(k_1,k_2,k_3,k_4) &=&
 W^S(k_2,k_1,k_3,k_4) =
 W^S(k_1,k_2,k_4,k_3) \nonumber \\ 
 &=& W^S(k_3,k_4,k_1,k_2) \; , \\
 W^T(k_1,k_2,k_3,k_4) &=&
 - W^T(k_2,k_1,k_3,k_4) =
 - W^T(k_1,k_2,k_4,k_3) \nonumber \\ 
 &=& W^T(k_3,k_4,k_1,k_2) \; .
\end{eqnarray}

Finally, a spin-rotation invariant anomalous interaction with
three creation and one annihilation operators, or vice versa,
can be written as
\begin{eqnarray} \label{gamma31}
 \Gam^{(3+1)}[\psi,\psib] &=& \frac{1}{2} \sum_{k_i}
 \Big\{ X^{S}(k_1,k_2,k_3,k_4) 
 \sum_{\sg} 
 \psib_{k_1\sg} (\psib_{k_2\up} \psib_{k_3\down} - 
 \psib_{k_2\down} \psib_{k_3\up}) \psi_{k_4\sg} \nonumber \\
 && + \; X^{T}(k_1,k_2,k_3,k_4)  
 \Big[ \sum_{\sg} \eps_{\sg}
 \psib_{k_1\sg} (\psib_{k_2\up} \psib_{k_3\down} + 
 \psib_{k_2\down} \psib_{k_3\up}) \psi_{k_4\sg} \nonumber \\
 && \quad + \; 
 2(\psib_{k_1\up}\psib_{k_2\down}\psib_{k_3\down}\psi_{k_4\down}
 - \psib_{k_1\down}\psib_{k_2\up}\psib_{k_3\up}\psi_{k_4\up})
 \Big] + {\rm conj.} \Big\} \; ,
\end{eqnarray}
where $\eps_{\up} = 1$ and $\eps_{\down} = -1$.
The terms with the coefficients $X^S$ and $X^T$ are separately
spin-rotation invariant.
Time reversal invariance yields the relation
\begin{equation}
 X^{S,T*}(k_1,k_2,k_3,k_4) = 
 X^{S,T}(-k_1,-k_2,-k_3,-k_4) 
\end{equation}
for a real choice of the order parameter.
Invariance under particle exchange yields
\begin{eqnarray}
 X^S(k_1,k_2,k_3,k_4) &=&
 \phantom- X^S(k_1,k_3,k_2,k_4) \; , \nonumber \\ 
 X^T(k_1,k_2,k_3,k_4) &=&
 - X^T(k_1,k_3,k_2,k_4) \; .
\end{eqnarray}

To express the two-particle interaction in terms of Nambu fields,
it is convenient to collect the 16 components of the Nambu vertex
$\Gam_{s_1s_2s_3s_4}^{(4)}$ in a $4 \times 4$ matrix
\begin{equation} \label{vertexmatrixdef}
 \bGam^{(4)} = \left( \begin{array}{cccc}
 \Gam_{++++}^{(4)} & \Gam_{++-+}^{(4)} & 
 \Gam_{+-++}^{(4)} & \Gam_{+--+}^{(4)} \\[2mm]
 \Gam_{+++-}^{(4)} & \Gam_{++--}^{(4)} & 
 \Gam_{+-+-}^{(4)} & \Gam_{+---}^{(4)} \\[2mm]
 \Gam_{-+++}^{(4)} & \Gam_{-+-+}^{(4)} & 
 \Gam_{--++}^{(4)} & \Gam_{---+}^{(4)} \\[2mm]
 \Gam_{-++-}^{(4)} & \Gam_{-+--}^{(4)} & 
 \Gam_{--+-}^{(4)} & \Gam_{----}^{(4)} 
 \end{array} \right) \quad .
\end{equation}
Note that the rows in this matrix are labelled by $s_1$ and $s_4$,
while columns are labelled by $s_2$ and $s_3$.
With this assignment the Bethe-Salpeter equation yielding the
exact Nambu vertex in reduced (mean-field) models can be written
as a matrix equation.
Translating the spin-rotation invariant structure of the various 
interaction terms described above to the Nambu representation,
one obtains a Nambu vertex of the following form
\begin{eqnarray} \label{vertexmatrix}
 \bGam^{(4)}(k_1,k_2,k_3,k_4) = \hskip 10.5cm  \nonumber \\[2mm]
 \left( \! \begin{array}{cccc}
 V^T(k_1,k_2,k_3,k_4) & X(k_1,k_2,k_3,k_4) & 
 X^*(k^*_4,k^*_3,k^*_2,k^*_1) & \!\! -V(k_1,\bar k_3,\bar k_2,k_4) \\[2mm]
 \!\! -X(k_1,k_2,k_4,k_3) & W(k_1,k_2,k_3,k_4) & 
 V(k_1,\bar k_4, \bar k_2,k_3) & X^*(k_4,k_3,k_1,k_2) \\[2mm]
 \!\! -X^*(k^*_3,k^*_4,k^*_2,k^*_1) & V^*(k_1,\bar k_4,\bar k_2,k_3) &
 W^*(k^*_4,k^*_3,k^*_2,k^*_1) & X(k^*_2,k^*_1,k^*_3,k^*_4) \\[2mm]
 \!\! -V^*(k_1,\bar k_3,\bar k_2,k_4) & \!\! -X^*(k_4,k_3,k_2,k_1) & 
 \!\! -X(k^*_1,k^*_2,k^*_3,k^*_4) & V^{T*}(k_1,k_2,k_3,k_4) 
 \end{array} \!\! \right) \nonumber \\[2mm]
\end{eqnarray}
where $\bar k = -k$. The matrix elements $W$ and $X$ are related to
the anomalous (4+0) and (3+1) interactions, respectively:
\begin{eqnarray}
 W(k_1,k_2,k_3,k_4) &=& 
 W^S(k_1,-k_4,-k_3,k_2) - W^S(k_1,-k_3,-k_4,k_2)
 \nonumber \\ 
 &+&
 W^T(k_1,-k_4,-k_3,k_2) - W^T(k_1,-k_3,-k_4,k_2)
 \nonumber \\ 
 &+&
 2 W^T(k_1,k_2,-k_3,-k_4) \; , \\[2mm]
 X(k_1,k_2,k_3,k_4) &=& 
 X^S(k_1,k_2,-k_3,k_4) - X^S(k_2,k_1,-k_3,k_4)
 \nonumber \\ 
 &+&
 X^T(k_1,k_2,-k_3,k_4) - X^T(k_2,k_1,-k_3,k_4)
 \nonumber \\ 
 &+&
 2 X^T(-k_3,k_2,k_1,k_4) \; .
\end{eqnarray}
The functions $W$ and $X$ obey the following relations under
exchange of variables:
\begin{eqnarray}
 W(k_1,k_2,k_3,k_4) &=& 
 -  W(k_2,k_1,k_3,k_4) = - W(k_1,k_2,k_4,k_3) \nonumber \\
 &=& \phantom- W(-k_4,-k_3,-k_2,-k_1) \; , \\[2mm]
 X(k_1,k_2,k_3,k_4) &=& - X(k_2,k_1,k_3,k_4) \; .
\end{eqnarray}

In summary, by fully exploiting the available symmetries, in 
particular spin-rotation invariance, the number of independent
functions parametrizing the Nambu vertex could be reduced 
substantially compared to what one gets by exploiting only the
conservation of the $z$-component of the spin.\cite{Gersch08}

\section{Reduced BCS plus forward scattering model}

In this section we consider a model with reduced interactions in the
Cooper and forward scattering channels. The truncated flow equations
described in {\S}2 yield the {\em exact} flow for that model.
By analyzing the flow we gain insight into the role of the various 
anomalous interactions and their singularities at and below the 
critical scale for superconductivity. 

The model is defined by the following action
\begin{equation} \label{redmodel}
 S[\psi,\psib] = S_0[\psi,\psib] +
 V_{sc}[\psi,\psib] + V_{c}[\psi,\psib] + 
 V_{s}[\psi,\psib] \; ,
\end{equation}
consisting of a quadratic and three interaction terms.
\begin{equation}
 S_0[\psi,\psib] = 
 \sum_{k,\sg} (-ik_0 + \xi_{\bk}) \psib_{k\sg} \psi_{k\sg} +
 \sum_k \left( \Delta^{(0)}(k) \psib_{k\up} \psib_{-k\down} +
 \Delta^{(0)*}(k) \psi_{-k\down} \psi_{k\up} \right)
\end{equation}
contains the kinetic energy and an external pairing field 
$\Delta^{(0)}$, while $V_{sc}$, $V_{c}$ and $V_{s}$ are 
reduced interactions in the Cooper (superconducting), forward 
charge, and forward spin (magnetic) channel, respectively:
\begin{eqnarray}
 V_{sc}[\psi,\psib] &=& 
 \sum_{k_1,k_2} V^{(0)}(k_1,k_2) \,
 \psib_{k_1\up} \psib_{-k_1\down} \psi_{-k_2\down} \psi_{k_2\up}
 \; , \\[2mm]
 V_{c}[\psi,\psib] &=&   
 \frac{1}{2} \sum_{k_1,k_2} \sum_{\sg_1,\sg_2} 
 F^{(0)}_c(k_1,k_2) \,
 \psib_{k_1\sg_1} \psib_{k_2\sg_2} \psi_{k_2\sg_2} \psi_{k_1\sg_1}
 \; , \\[2mm]
 V_{s}[\psi,\psib] &=&   
 \frac{1}{2} \sum_{k_1,k_2} \sum_{\sg_1,\sg'_1} \sum_{\sg_2,\sg'_2} 
 F^{(0)}_s(k_1,k_2) \left( \vec\tau_{\sg_1\sg'_1} \cdot 
 \vec\tau_{\sg_2\sg'_2} \right) \,
 \psib_{k_1\sg_1} \psib_{k_2\sg_2} \psi_{k_2\sg'_2} \psi_{k_1\sg'_1}
 \; , \hskip 1cm
\end{eqnarray}
where $\vec\tau = (\tau^x,\tau^y,\tau^z)$ is the vector formed by
the three Pauli matrices. The coupling functions $V^{(0)}(k_1,k_2)$,
$F^{(0)}_c(k_1,k_2)$ and $F^{(0)}_s(k_1,k_2)$ are arbitrary 
functions of $k_1$ and $k_2$.
The interaction terms correspond to a spin-rotation invariant 
normal vertex of the form (\ref{gamma22}) with 
\begin{eqnarray} \label{Vred}
 V(k_1,k_2,k_3,k_4) &=& 
 V^{(0)}(k_1,k_4) \delta_{k_1,-k_2} \delta_{k_3,-k_4} +
 F^{(0)}_c(k_1,k_2) \delta_{k_1,k_4} \delta_{k_2,k_3} \nonumber \\
 &-& F^{(0)}_s(k_1,k_2) (\delta_{k_1,k_4} \delta_{k_2,k_3} +
 2 \delta_{k_1,k_3} \delta_{k_2,k_4}) \; .
\end{eqnarray}
The pairing field $\Delta^{(0)}$ breaks the $U(1)$ charge symmetry
explicitly. Spontaneous symmetry breaking is obtained in the
limit $\Delta^{(0)} \to 0$.
The model (\ref{redmodel}) is a generalization of the
reduced BCS model, where only $V_{sc}$ contributes, whose 
flow was discussed extensively in Ref.~\citen{Salmhofer04}.
A special version of the model (\ref{redmodel}) with $V_{s} = 0$ 
was solved  by a summation of all contributing Feynman diagrams 
in Ref.~\citen{Gersch07}.

\subsection{Exact integral equations and Ward identity}

Due to the restricted momentum dependence of the interaction
terms, all contributions to $\Gam^{(2)\Lam}$ and $\Gam^{(4)\Lam}$ 
discarded in the truncation described in {\S}2 vanish in the 
thermodynamic limit. 
This follows from a straightforward generalization of the 
arguments given in Ref.~\citen{Salmhofer04}.
The constrained momentum dependence of the interactions
leads to the following constraints on the external momenta
of the various one-loop contributions to the flow of the
Nambu vertex $\Gam^{(4)\Lam}_{s_1s_2s_3s_4}(k_1,k_2,k_3,k_4)$,
Eq.~(\ref{floweq_vertex}):
\begin{eqnarray}
 \Pi^{PH,d}_{s_1s_2s_3s_4}(k_1,k_2,k_3,k_4) &\propto&
 \delta_{k_1,k_4} \delta_{k_2,k_3} \; , \nonumber \\
 \Pi^{PH,cr}_{s_1s_2s_3s_4}(k_1,k_2,k_3,k_4) &\propto&
 \delta_{k_1,k_3} \delta_{k_2,k_4} \; , \nonumber \\
 \Pi^{PP}_{s_1s_2s_3s_4}(k_1,k_2,k_3,k_4) &\propto&
 \delta_{k_1,-k_2} \delta_{k_3,-k_4} \; .
\end{eqnarray}

It is sufficient to consider the flow equation for the 
Nambu vertex with $k_1 = k_4$ and $k_2 = k_3$, since the
non-vanishing matrix elements for other choices of momenta
follow from symmetries.
Choosing $k_1 = k_4 =: k$ and $k_2 = k_3 =: k'$, only the
direct particle-hole term contributes and the flow
equation simplifies to
\begin{eqnarray}
 \frac{d}{d\Lam}
 \Gam^{(4)\Lam}_{s_1s_2s_3s_4}(k,k',k',k) 
 =& \hskip -2cm 
 \Pi^{PH,d}_{s_1s_2s_3s_4}(k,k',k',k) =
 \nonumber \\
 \sum_{p,s'_i} \frac{d}{d\Lam}
 [G^{\Lam}_{s'_1s'_2}(p) G^{\Lam}_{s'_3s'_4}(p)] & 
 \Gam^{(4)\Lam}_{s_1s'_2s'_3s_4}(k,p,p,k)
 \Gam^{(4)\Lam}_{s'_4s_2s_3s'_1}(p,k',k',p) \; .
\end{eqnarray}
This differential equation is equivalent to the 
Bethe-Salpeter-like integral equation
\begin{eqnarray} \label{exactvertex}
 \hskip -5mm
 \Gam^{(4)\Lam}_{s_1s_2s_3s_4}(k,k',k',k) 
 &=& \Gam^{(4)\Lam_0}_{s_1s_2s_3s_4}(k,k',k',k)
 \nonumber \\[2mm]
 + \sum_{p,s'_i} && \hskip -5mm
 G^{\Lam}_{s'_1s'_2}(p) G^{\Lam}_{s'_3s'_4}(p) 
 \Gam^{(4)\Lam_0}_{s_1s'_2s'_3s_4}(k,p,p,k)
 \Gam^{(4)\Lam}_{s'_4s_2s_3s'_1}(p,k',k',p)
\end{eqnarray}
summing up all Nambu particle-hole ladder diagrams.
Note that $G^{\Lam_0} = 0$. 
Inserting this implicit solution for $\Gam^{(4)\Lam}$
into the flow equation (\ref{floweq_selfen}) for the
self-energy, and using the relation 
$\bS^{\Lam} = \dot\bG^{\Lam} + 
 \bG^{\Lam} \dot\bSg^{\Lam} \bG^{\Lam}$, the flow of
the self-energy can be integrated, yielding
\begin{equation} \label{exactselfen}
 \Sg^{\Lam}_{s_1s_2}(k) - \Sg^{\Lam_0}_{s_1s_2}(k) =
 - \sum_{k'} \sum_{s_3,s_4} G^{\Lam}_{s_4s_3}(k')
 \Gam^{(4)\Lam_0}_{s_1s_3s_4s_2}(k,k',k',k) \; ,
\end{equation}
with the bare Nambu vertex on the right hand side.
This is just the familiar mean-field equation for the 
self-energy, which is exact for the reduced model 
(\ref{redmodel}).

The exact solution determined by Eqs.~(\ref{exactvertex})
and (\ref{exactselfen}) fulfils the Ward identity following 
from global charge conservation,\cite{Salmhofer04}
\begin{eqnarray} \label{ward}
 \Delta^{\Lam}(k) - \Delta^{(0)}(k) &=&
 \sum_{k'} \sum_{s,s'} \left[
 \Delta^{(0)}(k') G^{\Lam}_{s+}(k') G^{\Lam}_{-s'}(k') -
 \Delta^{(0)*}(k') G^{\Lam}_{s-}(k') G^{\Lam}_{+s'}(k') 
 \right] \nonumber \\
 && \times \Gam^{(4)\Lam}_{+s's-}(k,k',k',k) \; ,  
\end{eqnarray}
which connects the anomalous self-energy 
$\Sg^{\Lam}_{+-}(k) = \Delta^{\Lam}(k)$ with the
two-particle vertex.
The Ward identity implies that some components of the
Nambu vertex diverge in case of spontaneous symmetry breaking 
($\Delta^{\Lam}$ finite for $\Delta^{(0)} \to 0$).

\subsection{Explicit solution and 
singularities of Nambu vertex}

An explicit solution of the Bethe-Salpeter equation
(\ref{exactvertex}) for the Nambu vertex can be obtained
for the case of separable momentum dependences of the
interaction terms in the reduced model.
In this special case the singularities of the various
components of the Nambu vertex become particularly
transparent.

Using the $4 \times 4$ matrix representation of the
Nambu vertex, Eq.~(\ref{vertexmatrixdef}), one can write
the integral equation (\ref{exactvertex}) in matrix form
as
\begin{equation}
 \bGam^{(4)\Lam}(k,k') =
 \bGam^{(4)\Lam_0}(k,k') + 
 \sum_p \bGam^{(4)\Lam_0}(k,p) \, \bL^{\Lam}(p) \, 
 \bGam^{(4)\Lam}(p,k') \; ,
\end{equation}
where $\bGam^{(4)\Lam}(k,k') = \bGam^{(4)\Lam}(k,k',k',k)$ and
$\bL^{\Lam}$ is the $4 \times 4$ matrix defined by
\begin{equation} \label{pairmatrix}
 \bL^{\Lam} = \left( \begin{array}{cccc}
 G^{\Lam}_{++} G^{\Lam}_{++} & G^{\Lam}_{-+} G^{\Lam}_{++} & 
 G^{\Lam}_{++} G^{\Lam}_{+-} & G^{\Lam}_{-+} G^{\Lam}_{+-} \\[2mm]
 G^{\Lam}_{++} G^{\Lam}_{-+} & G^{\Lam}_{-+} G^{\Lam}_{-+} & 
 G^{\Lam}_{++} G^{\Lam}_{--} & G^{\Lam}_{-+} G^{\Lam}_{--} \\[2mm]
 G^{\Lam}_{+-} G^{\Lam}_{++} & G^{\Lam}_{--} G^{\Lam}_{++} & 
 G^{\Lam}_{+-} G^{\Lam}_{+-} & G^{\Lam}_{--} G^{\Lam}_{+-} \\[2mm]
 G^{\Lam}_{+-} G^{\Lam}_{-+} & G^{\Lam}_{--} G^{\Lam}_{-+} & 
 G^{\Lam}_{+-} G^{\Lam}_{--} & G^{\Lam}_{--} G^{\Lam}_{--}
 \end{array} \right) \quad .
\end{equation}
The bare Nambu vertex $\bGam^{(4)\Lam_0}(k,k')$
corresponding to the reduced interaction in Eq.~(\ref{Vred})
has the form
\begin{equation} \label{barevertexred}
 \bGam^{(4)\Lam_0}(k,k') = \left( \begin{array}{cccc}
 F^{(0)}_{1}(k,k') & 0 & 
 0 & F^{(0)}_{2}(k,k') \\[2mm]
 0 & 0 & V^{(0)}(k,k') & 0 \\[2mm]
 0 & V^{(0)*}(k,k') & 0 & 0 \\[2mm]
 F^{(0)*}_{2}(k,k') & 0 & 
 0 & F^{(0)*}_{1}(k,k')
 \end{array} \right) \; ,
\end{equation}
where
\begin{eqnarray}
 F^{(0)}_{1}(k,k') &=& 
 F^{(0)}_c(k,k') + F^{(0)}_s(k,k') \; , \nonumber \\ 
 F^{(0)}_{2}(k,k') &=& 
 F^{(0)}_s(k,-k') - F^{(0)}_c(k,-k') \; .
\end{eqnarray}
The structure of the full Nambu vertex 
$\bGam^{(4)\Lam}(k,k')$ is obtained by specializing
the general form Eq.~(\ref{vertexmatrix}) to the case 
$k_1 = k_4 = k$ and $k_2 = k_3 = k'$ as
\begin{equation} \label{fullvertexred}
 \bGam^{(4)\Lam}(k,k') = \left( \begin{array}{cccc}
 F^{\Lam}_{1}(k,k') & X^{\Lam}(k,k') & 
 X^{\Lam *}(k^*,{k'}^*) & F^{\Lam}_{2}(k,k') \\[2mm]
 X^{\Lam}(k',k) & W^{\Lam}(k,k') & 
 V^{\Lam}(k,k') & -X^{\Lam *}(k',k) \\[2mm]
 X^{\Lam *}({k'}^*,k^*) & V^{\Lam *}(k,k') & 
 W^{\Lam *}(k^*,{k'}^*) & -X^{\Lam}({k'}^*,k^*) \\[2mm]
 F^{\Lam *}_{2}(k,k') & -X^{\Lam *}(k,k') & 
 -X^{\Lam}(k^*,{k'}^*) & F^{\Lam *}_{1}(k,k')
 \end{array} \right) \; .
\end{equation}

In the following we assume that the bare coupling functions
$V^{(0)}$, $F^{(0)}_c$, $F^{(0)}_s$, and the external pairing 
field $\Delta^{(0)}$ are real and frequency independent.
Then $\bGam^{(4)\Lam}(k,k')$ and $\Delta^{\Lam}(k)$ are 
also real and frequency independent, such that the complex 
conjugation operations in Eq.~(\ref{fullvertexred}) can be 
omitted.
The matrix products in the Bethe-Salpeter equation 
(\ref{exactvertex}) can then be simplified considerably by 
employing the following orthogonal transformation
\begin{equation} \label{orthotrafo}
 \bU = \frac{1}{\sqrt{2}} \left( \begin{array}{rrcc}
  1 &  0 & 0 & 1 \\[2mm]
  0 &  1 & 1 & 0 \\[2mm]
  0 & -1 & 1 & 0 \\[2mm]
 -1 &  0 & 0 & 1
 \end{array} \right) \; .
\end{equation}
The transformed vertex 
$\tilde\bGam^{(4)\Lam} = \bU \bGam^{(4)\Lam} \bU^T$ has the form
\begin{equation} \label{fullvertexredtrans}
 \tilde\bGam^{(4)\Lam}(k,k') = \left( \begin{array}{cccc}
 F^{\Lam}_{+}(k,k') & 0 & 0 & 0 \\[2mm]
 0 & A^{\Lam}(k,k') & 0 & -2X^{\Lam}(k',k) \\[2mm]
 0 & 0 & - \Phi^{\Lam}(k,k') & 0 \\[2mm]
 0 & -2X^{\Lam}(k,k') & 0 & F^{\Lam}_{-}(k,k')
 \end{array} \right) \; ,
\end{equation}
with the ''amplitude'' ($A$) and ''phase'' ($\Phi$) components
\begin{eqnarray}
 A^{\Lam}(k,k') &=& 
 V^{\Lam}(k,k') + W^{\Lam}(k,k') \; , \nonumber \\
 \Phi^{\Lam}(k,k') &=& 
 V^{\Lam}(k,k') - W^{\Lam}(k,k') \; ,
\end{eqnarray}
and the linear combinations in the forward scattering channel
\begin{equation}
 F^{\Lam}_{\pm}(k,k') = 
 F^{\Lam}_1(k,k') \pm F^{\Lam}_2(k,k')  \; .
\end{equation}
The transformed bare vertex $\tilde\bGam^{(4)\Lam_0}(k,k')$
has only diagonal entries:
\begin{equation} \label{barevertexredtrans}
 \tilde\bGam^{(4)\Lam_0}(k,k') = \left( \begin{array}{cccc}
 F^{(0)}_{+}(k,k') & 0 & 0 & 0 \\[2mm]
 0 & V^{(0)}(k,k') & 0 & 0 \\[2mm]
 0 & 0 & - V^{(0)}(k,k') & 0 \\[2mm]
 0 & 0 & 0 & F^{(0)}_{-}(k,k')
 \end{array} \right) \; .
\end{equation}
The frequency summed matrix 
$\bL^{\Lam}(\bp) = \sum_{p_0} \bL^{\Lam}(p)$ 
also transforms to a simpler block matrix structure
\begin{equation} \label{pairmatrixtrans}
 \tilde\bL^{\Lam}(\bp) = \left( \begin{array}{cccc}
 L^{\Lam}_{f_+}(\bp) & 0 & 0 & 0 \\[2mm]
 0 & L^{\Lam}_a(\bp) & 0 & L^{\Lam}_x(\bp) \\[2mm]
 0 & 0 & L^{\Lam}_{\phi}(\bp) & 0 \\[2mm]
 0 & L^{\Lam}_x(\bp) & 0 & L^{\Lam}_{f_-}(\bp)
 \end{array} \right) \; ,
\end{equation}
where
\begin{eqnarray} 
 L^{\Lam}_a(\bp) &=& \sum_{p_0} 
 [G^{\Lam}_{+-}(p)]^2 + G^{\Lam}_{++}(p) G^{\Lam}_{--}(p) 
 \; , \nonumber \\ 
 L^{\Lam}_{\phi}(\bp) &=& \sum_{p_0} 
 [G^{\Lam}_{+-}(p)]^2 - G^{\Lam}_{++}(p) G^{\Lam}_{--}(p)
 \; , \nonumber \\
 L^{\Lam}_x(\bp) &=& -2 \sum_{p_0} 
 G^{\Lam}_{++}(p) G^{\Lam}_{+-}(p)
 \; , \nonumber \\
 L^{\Lam}_{f_+}(\bp) &=& \sum_{p_0} 
 [G^{\Lam}_{++}(p)]^2 + [G^{\Lam}_{+-}(p)]^2
 \; , \nonumber \\
 L^{\Lam}_{f_-}(\bp) &=& \sum_{p_0} 
 [G^{\Lam}_{++}(p)]^2 - [G^{\Lam}_{+-}(p)]^2 \; .
\end{eqnarray}

For an explicit solution, we now assume that the momentum
dependences of the bare interactions factorize:
\begin{eqnarray} \label{baresepintact}
 V^{(0)}(k,k') &=& 
 g^{(0)} f_{sc}(\bk) f_{sc}(\bk') \; , \nonumber \\[2mm]
 F^{(0)}_c(k,k') &=& 
 g^{(0)}_{c} f_{c}(\bk) f_{c}(\bk') \; , \nonumber \\[2mm]
 F^{(0)}_s(k,k') &=& 
 g^{(0)}_{s} f_{s}(\bk) f_{s}(\bk') \; ,
\end{eqnarray}
where $f_{sc}(\bk)$, $f_{c}(\bk)$, and $f_{s}(\bk)$ are
arbitrary reflection invariant form factors.
Then the momentum dependences of the flowing interactions 
also factorize in the form
\begin{eqnarray}
 V^{\Lam}(k,k') &=& 
 g^{\Lam} f_{sc}(\bk) f_{sc}(\bk') \; , \nonumber \\[2mm]
 F^{\Lam}_{-}(k,k') &=& 
 2g^{\Lam}_{c} f_{c}(\bk) f_{c}(\bk') \; , \nonumber \\[2mm]
 F^{\Lam}_{+}(k,k') &=& 
 2g^{\Lam}_{s} f_{s}(\bk) f_{s}(\bk') \; , \nonumber \\[2mm]
 W^{\Lam}(k,k') &=& 
 g^{\Lam}_{w} f_{sc}(\bk) f_{sc}(\bk') \; , \nonumber \\[2mm]
 X^{\Lam}(k,k') &=& 
 g^{\Lam}_{x} f_{c}(\bk) f_{sc}(\bk') \; ,
\end{eqnarray}
and also the linear combinations $V^{\Lam} \pm W^{\Lam}$,
\begin{eqnarray}
 A^{\Lam}(k,k') &=& 
 g^{\Lam}_{a} f_{sc}(\bk) f_{sc}(\bk') \; , \nonumber \\[2mm]
 \Phi^{\Lam}(k,k') &=& 
 g^{\Lam}_{\phi} f_{sc}(\bk) f_{sc}(\bk') \; ,
\end{eqnarray}
with $g^{\Lam}_{a} = g^{\Lam} + g^{\Lam}_{w}$ 
and  $g^{\Lam}_{\phi} = g^{\Lam} - g^{\Lam}_{w}$.

The momentum dependences on the right hand side of the 
integral equation (\ref{exactvertex}) now factorize,
and the momentum integration can be isolated in the 
following numbers:
\begin{eqnarray}
 l^{\Lam}_a &=& 
 \sum_{\bp} L^{\Lam}_a(\bp) f_{sc}^2(\bp)
 \; , \nonumber \\
 l^{\Lam}_{\phi} &=& 
 \sum_{\bp} L^{\Lam}_{\phi}(\bp) f_{sc}^2(\bp)
 \; , \nonumber \\
 l^{\Lam}_x &=& 
 \sum_{\bp} L^{\Lam}_x(\bp) f_c(\bp) f_{sc}(\bp)
 \; , \nonumber \\
 l^{\Lam}_c &=& 
 \sum_{\bp} L^{\Lam}_{f_-}(\bp) f_c^2(\bp)
 \; , \nonumber \\
 l^{\Lam}_s &=& 
 \sum_{\bp} L^{\Lam}_{f_+}(\bp) f_s^2(\bp)  \; .
\end{eqnarray}
Inverting the resulting matrix equation, one obtains the
explicit solution for the flowing couplings:
\begin{eqnarray}
 g^{\Lam}_a &=& 
 \frac{(g^{(0)}_c)^{-1} - 2l^{\Lam}_c}{2d^{\Lam}} 
 \nonumber \\[2mm]
 g^{\Lam}_{\phi} &=& 
 \frac{1}{(g^{(0)})^{-1} + l^{\Lam}_{\phi}}
 \nonumber \\[2mm]
 g^{\Lam}_x &=& 
 - \frac{l^{\Lam}_x}{2d^{\Lam}} 
 \nonumber \\[2mm]
 g^{\Lam}_c &=& 
 \frac{(g^{(0)})^{-1} - l^{\Lam}_a}{2d^{\Lam}} 
 \nonumber \\[2mm]
 g^{\Lam}_s &=& 
 \frac{1}{(g^{(0)}_s)^{-1} - 2l^{\Lam}_s}
 \; ,
\end{eqnarray}
where
\begin{equation}
 d^{\Lam} = \left[ (g^{(0)})^{-1} - l^{\Lam}_a \right]
 \left[ (2g^{(0)}_c)^{-1} - l^{\Lam}_c \right] - 
 \left( l^{\Lam}_x \right)^2 \; .
\end{equation}
Note that $g^{\Lam}_s$ and $g^{\Lam}_{\phi}$ are coupled to
other interactions only indirectly via the propagators,
while $g^{\Lam}_a$, $g^{\Lam}_c$, and $g^{\Lam}_x$ are 
coupled also directly.

The mean-field equation for the self-energy (\ref{exactselfen})
implies that the interaction driven part of the gap function,
$\Delta^{\Lam}(k) - \Delta^{(0)}(k)$, adopts the momentum 
dependence of the form factor for $V^{(0)}(k,k')$. 
Assuming also $\Delta^{(0)}(k) = \delta^{(0)} f_{sc}(\bk)$, 
we can write
\begin{equation}
 \Delta^{\Lam}(k) = \delta^{\Lam} f_{sc}(\bk) \; .
\end{equation}

We now discuss the behavior of the various components of the
Nambu vertex as a function of the flow parameter $\Lam$,
especially the singularities near the critical scale $\Lam_c$
for spontaneous $U(1)$-symmetry breaking, at which the vertex 
diverges for $\Delta^{(0)} \to 0$. 
We consider the usual case where superconductivity is the only 
instability of the system, that is, no instabilities driven by 
forward scattering occur ($g^{\Lam}_c$ and $g^{\Lam}_s$ remain 
finite for all $\Lam$).

For $\Lam > \Lam_c$ and $\Delta^{(0)} = 0$ the anomalous 
propagator $F(k)$ vanishes, such that 
$l^{\Lam}_a = - l^{\Lam}_{\phi}$, $l^{\Lam}_x = 0$, and
$l^{\Lam}_c = l^{\Lam}_s$.
The anomalous components of the Nambu vertex $W^{\Lam}$
and $X^{\Lam}$ also vanish, and
\begin{equation}
 V^{\Lam}(k,k') = 
 \frac{f_{sc}(\bk) f_{sc}(\bk')}{(g^{(0)})^{-1} - l^{\Lam}_a}
 \; .
\end{equation}
The critical scale $\Lam_c$ is the scale where
\begin{equation}
 l^{\Lam_c}_a = 
 - \sum_p G^{\Lam_c}(p) G^{\Lam_c}(-p) f_{sc}^2(\bp) =
 (g^{(0)})^{-1} \; ,
\end{equation}
such that $V(k,k')$ diverges.

For $\Lam < \Lam_c$ and/or $\Delta^{(0)} \neq 0$, anomalous
components appear. 
Using the relation
\begin{equation}
 F(k) = \Delta(k) [ G(k) G(-k) + F(k) F^*(-k) ]  \; ,
\end{equation}
one can write the gap equation contained in 
Eq.~(\ref{exactselfen}) in the form
\begin{equation}
 \delta^{\Lam} - \delta^{(0)} = 
 - g^{(0)} \delta^{\Lam} l^{\Lam}_{\phi} \; .
\end{equation}
Inserting this into the solution for $\Phi^{\Lam}(k,k')$,
one finds
\begin{equation} \label{goldstone}
 \Phi^{\Lam}(k,k') = 
 \frac{\delta^{\Lam}}{\delta^{(0)}} V^{(0)}(k,k')
 \; .
\end{equation}
Note that the forward scattering interactions and the
anomalous (3+1)-components of the vertex do not affect
this result. For $\Delta^{(0)} \to 0$, one finds
$\Phi^{\Lam}(k,k') \to \infty$ for any $\Lam < \Lam_c$.
This is the divergence required by the Ward identity 
(\ref{ward}) following from global charge conservation,
and is associated with a massless Goldstone boson in 
the symmetry-broken state.

In contrast to the phase component $\Phi^{\Lam}$ of the 
vertex, the amplitude component $A^{\Lam}$ is regularized
by the gap below the critical scale. 
Slightly below $\Lam_c$, it behaves as
\begin{equation}
 A^{\Lam}(k,k') \propto
 \frac{1}{\frac{\delta^{(0)}}{\delta^{\Lam}} 
 + \cO[(\delta^{\Lam})^2]} \; .
\end{equation}
For $\Delta^{(0)} \to 0$, this component is therefore of
order $(\delta^{\Lam})^{-2}$ slightly below $\Lam_c$.
The anomalous (3+1)-interaction behaves as
\begin{equation}
 X^{\Lam}(k,k') \propto
 \frac{\delta^{\Lam}}{\frac{\delta^{(0)}}{\delta^{\Lam}} 
 + \cO[(\delta^{\Lam})^2]} \; .
\end{equation}
For $\Delta^{(0)} \to 0$, it is therefore of order 
$(\delta^{\Lam})^{-1}$ slightly below $\Lam_c$.

We finally illustrate the above results by plotting the 
flow of the self-energy and the vertex components for a
specific choice of model parameters and at zero 
temperature.
The kinetic energy enters only via the density of states, 
which we choose as a constant
$D(\xi) = 1$ with $|\xi| \leq 1$. 
All form factors in the interaction terms are chosen as
unity, $f_{sc}(\bk) = f_c(\bk) = f_s(\bk) = 1$,
and the bare couplings as
$g^{(0)} = -0.5$, $g^{(0)}_c = 0.2$, and $g^{(0)}_s = 0$.
The qualitative features of the flow do not depend on the
size of the couplings. The spin coupling has been chosen
zero because the spin channel is decoupled from the rest.
The flow is computed for a sharp cutoff acting on the bare 
kinetic energy, such that $|\xi| > \Lam$.
Since all bare interactions are momentum independent, also
the flowing self-energy and the various components of the
flowing Nambu vertex are momentum independent.

In Fig.~3 we show the flow of the gap $\delta^{\Lam}$ for 
various values of the initial gap $\delta^{(0)}$. Note the 
sharp onset of $\delta^{\Lam}$ at the critical scale $\Lam_c$ 
for $\delta^{(0)} = 0$. This singularity is obviously smeared 
out for $\delta^{(0)} > 0$.
\begin{figure}
\centerline{\includegraphics[width=8cm]{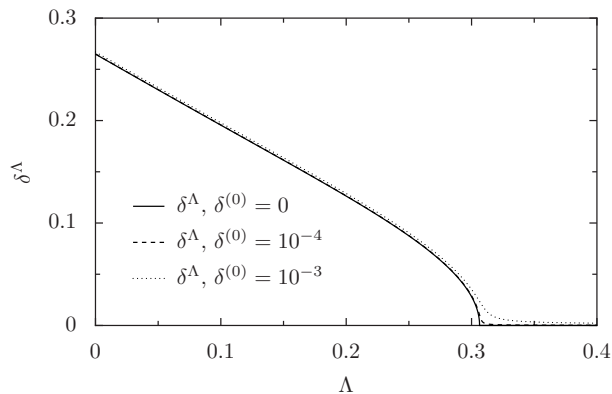}}
\caption{Flow of the superconducting gap $\delta^{\Lam}$ 
 for various values of the bare gap $\delta^{(0)}$.}
\end{figure}
The flow of the phase component of the Nambu vertex
$\Phi^{\Lam}$ is shown in the left panel of Fig.~4, and
its final value at $\Lam=0$ as a function of $\delta^{(0)}$ 
in the right panel. 
Note that $\Phi^{\Lam}$ has the same shape as $\delta^{\Lam}$, 
and that $\Phi^{\Lam=0}$ diverges as $1/\delta^{(0)}$
for small $\delta^{(0)}$, as described by 
Eq.~(\ref{goldstone}).
\begin{figure}
\centerline{\includegraphics[width=10cm]{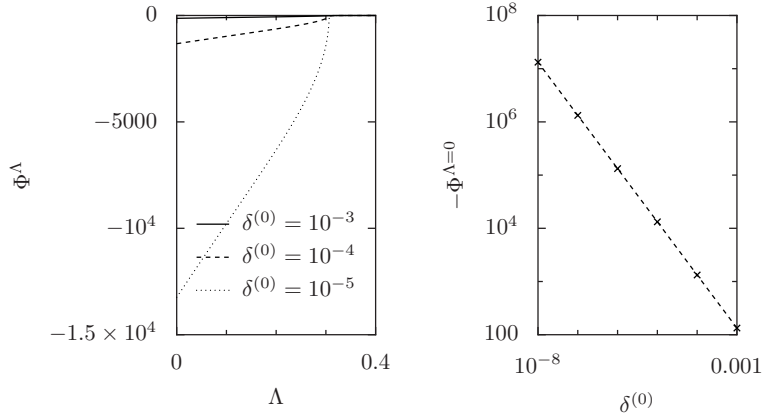}}
\caption{Left: Flow of the phase component of the Nambu
 vertex $\Phi^{\Lam}$ for various values of $\delta^{(0)}$. 
 Right: Phase component of the Nambu vertex at the end
 of the flow ($\Lam=0$) as a function of $\delta^{(0)}$.
 Note that $\Phi^{\Lam}$ is momentum independent due to the 
 choice of a constant form factor $f_{sc}(\bk) = 1$.}
\end{figure}
In Fig.~5 we plot the flow of the amplitude component
of the Nambu vertex $A^{\Lam}$ and of the anomalous
component related to (3+1)-interactions $X^{\Lam}$.
Both diverge at the critical scale $\Lam_c$ for
$\delta^{(0)} \to 0$, but decrease again for 
$\Lam < \Lam_c$. Note that $X^{\Lam}$ is much smaller
than $A^{\Lam}$ for any $\Lam$.
\begin{figure}
\centerline{\includegraphics[width=10cm]{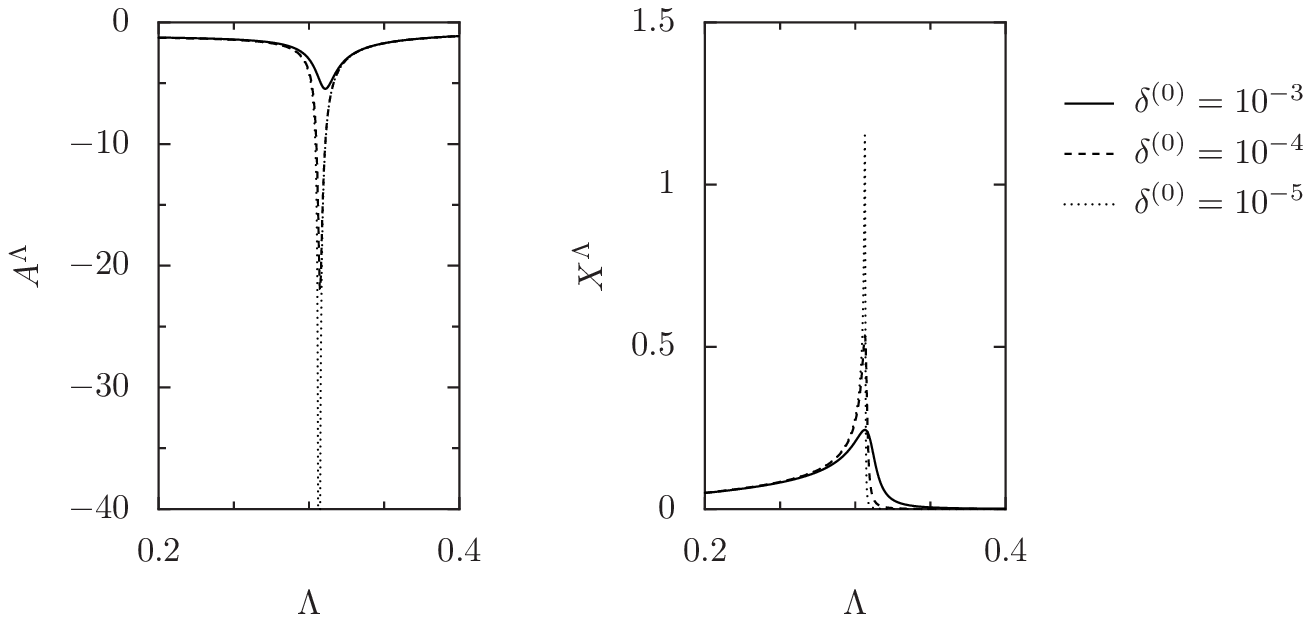}}
\caption{Flow of the amplitude component $A^{\Lam}$ (left) 
 of the Nambu vertex and of the anomalous component 
 $X^{\Lam}$ corresponding to (3+1)-interactions (right)
 for various values of $\delta^{(0)}$.
 Note that $A^{\Lam}$ and $X^{\Lam}$ are momentum independent 
 due to the choice $f_{sc}(\bk) = f_c(\bk) = 1$.}
\end{figure}

\section{Channel decomposition}

When solving the flow equation for the two-particle vertex
for a realistic (not mean field) model,
one faces the problem that rather complex momentum and 
frequency dependences are generated by the flow.
In a translation invariant system, the two-particle vertex 
depends on three independent momentum and frequency
variables.
In previous works the frequency dependence was usually
discarded, and the momentum dependence was crudely
discretized.\cite{Zanchi00,Halboth00,Honerkamp01,Gersch08}
That procedure can be justified by power counting, as long
as the flowing interactions remain weak and regular.
However, in case of spontaneous symmetry breaking the
two-particle vertex diverges in certain channels and
singular momentum and frequency dependences develop.

In this section we present a first step toward an efficient
parametrization of the two-particle vertex, which captures 
the singular momentum and frequency dependences associated 
with the flow into a superconducting phase.
The singular momentum dependence of the vertices originates 
from that of the loop integrals in the RG equations. 
Therefore, we decompose the vertices into several interaction 
channels, with the purpose of isolating the singular momentum 
dependence of each diagram. 
Distributing the diagrams among the channels according to their 
momentum dependence then leads to channel-decomposed flow
equations. 

For the normal vertices, we follow the approach by Husemann 
and Salmhofer \cite{Husemann09} and add several general 
two-particle interaction terms for the different channels 
to the bare interaction. The latter is not distributed among the 
channels to avoid ambiguities. The interaction channels then 
describe corrections to the bare interaction from integrating 
out modes in the functional integral. We make the ansatz
\begin{eqnarray}
 \Gam^{(2+2),\Lam}[\psi,\psib] &=& V[\psi,\psib] +
 \frac{1}{2} \sum_{k_i,\sg_i} \delta_{k_1+k_2,k_3+k_4}
 \nonumber \\
 &\times&
 \left\{ \left[ 
 V^{sc,\Lam}_{k_1,k_4}(k_1 + k_2) + 
 V^{c,\Lam}_{k_1,k_2}(k_3 - k_2) 
 \right]
 \delta_{\sg_1\sg_4} \delta_{\sg_2\sg_3} \right.
 \nonumber \\
 && + \left. V^{s,\Lam}_{k_1,k_2}(k_3 - k_2)
 \left( 2 \delta_{\sg_1\sg_3} \delta_{\sg_2\sg_4} -
 \delta_{\sg_1\sg_4} \delta_{\sg_2\sg_3}
 \right) \right\} \nonumber \\
 &\times& \psib_{k_1\sg_1} \psib_{k_2\sg_2} 
 \psi_{k_3\sg_3} \psi_{k_4\sg_4} \; ,
 \hskip 8mm
\end{eqnarray}
where $V[\psi,\psib]$ is the bare interaction and the superscripts 
{\it sc}, $c$, and $s$ refer to ''superconducting'', ''charge'', 
and ''spin'' channels, respectively.
The momentum (and frequency) argument in parentheses
is the total momentum of the interacting electrons 
for the superconducting channel and the momentum transfer
for the charge and spin channels.
These are the variables for which a singular dependence 
is expected. 
If the other momentum dependences remain regular, the above
ansatz provides a good starting point for a parametrization
of the various interaction channels as interactions mediated
by a boson exchange.

After symmetrization and translation to Nambu notation, one
of the Nambu components representing the (2+2)-interaction
reads
\begin{eqnarray} \label{channeldecom22}
 \Gam^{(4),\Lam}_{+-+-}(k_1,k_2,k_3,k_4) &=&
 V^{(0)}(k_1,-k_4,-k_2,k_3) + 
 V^{sc,\Lam}_{k_1,k_3}(k_3-k_2) + 
 V^{c,\Lam}_{k_1,-k_4}(k_1-k_3) \nonumber \\
 &-& 2 V^{s,\Lam}_{k_1,-k_4}(k_1+k_2) -
 V^{s,\Lam}_{k_1,-k_4}(k_1-k_3) \; ,
\end{eqnarray}
where $k_4 = k_1 + k_2 - k_3$.
Note that the total momentum of electrons in the 
superconducting channel has transformed to a momentum 
transfer in the Nambu representation, while the momentum
sum $k_1+k_2$ now appears in one of the spin-channel
terms.

The propagators in Eq.~(\ref{bubbles}) transport the
following momenta through the diagrams, which they 
depend on singularly:
\begin{eqnarray}
 \Pi^{PH,d}_{s_1s_2s_3s_4}(k_1,k_2,k_3,k_4) &:& \;
 \mbox{momentum transfer } k_3 - k_2 \; ,
 \nonumber \\
 \Pi^{PH,cr}_{s_1s_2s_3s_4}(k_1,k_2,k_3,k_4) &:& \;
 \mbox{momentum transfer } k_3 - k_1 \; ,
 \nonumber \\
 \Pi^{PP}_{s_1s_2s_3s_4}(k_1,k_2,k_3,k_4) &:& \;
 \mbox{total momentum } k_1 + k_2 \; .
 \nonumber
\end{eqnarray}
The three terms contributing to the flow of the two-particle
vertex in Eq.~(\ref{floweq_vertex}) are distributed among the 
various interaction channels according to their momentum 
dependence.
Hence we assign the direct Nambu particle-hole diagram
to the superconducting channel, such that
\begin{equation}
 \frac{d}{d\Lam} V^{sc,\Lam}_{k_1,k_3}(q) =
 \Pi^{PH,d}_{+-+-}(k_1,k_3-q,k_3,k_1-q) \; ,
\end{equation}
where the dependence on $q = k_3 - k_2$ is expected to be
singular in the superconducting phase.
The Nambu particle-particle diagram is assigned to the
spin channel,
\begin{equation}
 \frac{d}{d\Lam} V^{s,\Lam}_{k_1,k_3}(p) =
 \frac{1}{4} \Pi^{PP}_{+-+-}(k_1,p-k_1,k_3+p,-k_3) \; ,
\end{equation}
with $p = k_1+k_2$.
The crossed Nambu particle-hole diagram determines the flow
of the remaining two terms in Eq.~(\ref{channeldecom22}),
that is,
\begin{equation}
 \frac{d}{d\Lam} V^{c,\Lam}_{k_1,k_2}(q') -
 \frac{d}{d\Lam} V^{s,\Lam}_{k_1,k_2}(q') =
 - \Pi^{PH,cr}_{+-+-}(k_1,-k_2-q',k_1-q',-k_2) \; ,
\end{equation}
with $q' = k_3 - k_1$.
Using the above flow equation for $V^{s,\Lam}$, this yields
\begin{equation}
 \frac{d}{d\Lam} V^{c,\Lam}_{k_1,k_2}(q') =
 \frac{1}{4} \Pi^{PP}_{+-+-}(k_1,q'-k_1,k_2+q',-k_2) -
 \Pi^{PH,cr}_{+-+-}(k_1,-k_2-q',k_1-q',-k_2) \; .
\end{equation}
We have thus obtained flow equations for all three 
contributions $V^{sc,\Lam}$, $V^{c,\Lam}$, and $V^{s,\Lam}$
to the normal interaction.
These channel decomposed flow equations for the normal 
interaction are equivalent to those derived by Husemann and
Salmhofer \cite{Husemann09} in the $\psi$-representation.

For the channel decomposition of the anomalous vertices,
we profit from the symmetries described in {\S}~3.
For the functions $X^{S,T}(k_1,k_2,k_3,k_4)$ 
parametrizing $\Gam^{(3+1)}[\psi,\psib]$ in Eq.~(\ref{gamma31}),
we expect a singular dependence on the variable $k_2 + k_3$, 
which is the total momentum of the Cooper pair contained in
$\Gam^{(3+1)}[\psi,\psib]$. We therefore write
\begin{eqnarray}
 X^{S,\Lam}(k_1,k_2,k_3,k_4) &=& 
 X^{S,\Lam}_{k_1,k_2}(k_2+k_3) \; , \nonumber \\
 X^{T,\Lam}(k_1,k_2,k_3,k_4) &=& 
 X^{T,\Lam}_{k_1,k_2}(k_2+k_3) \; ,
\end{eqnarray}
where $k_4 = k_1 + k_2 + k_3$.
As a Nambu component representing the (3+1)-interaction we 
choose 
\begin{eqnarray}
 \Gam^{(4),\Lam}_{++-+}(k_1,k_2,k_3,k_4) &=&
 X^{\Lam}(k_1,k_2,k_3,k_4) \nonumber \\
 &=& X^{S,\Lam}_{k_1,k_2}(k_2-k_3) +
  X^{T,\Lam}_{k_1,k_2}(k_2-k_3) \nonumber \\
 &-& X^{S,\Lam}_{k_2,k_1}(k_1-k_3) -
  X^{T,\Lam}_{k_2,k_1}(k_1-k_3) +
  2X^{T,\Lam}_{-k_3,k_2}(k_1+k_2) \; , \hskip 8mm
\end{eqnarray}
where now $k_4 = k_1 + k_2 - k_3$.
Comparing the momentum dependences with those of the bubbles,
we assign
\begin{eqnarray}
 \frac{d}{d\Lam} \left[ X^{S,\Lam}_{k_1,k_2}(k_2-k_3) +
  X^{T,\Lam}_{k_1,k_2}(k_2-k_3) \right] &=&
 \Pi^{PH,d}_{++-+}(k_1,k_2,k_3,k_1+k_2-k_3) \; ,
 \label{xdirect} \\
 \frac{d}{d\Lam} \left[ X^{S,\Lam}_{k_2,k_1}(k_1-k_3) +
  X^{T,\Lam}_{k_2,k_1}(k_1-k_3) \right] &=&
 \Pi^{PH,cr}_{++-+}(k_1,k_2,k_3,k_1+k_2-k_3) \; ,
 \label{xcrossed} \\
 \frac{d}{d\Lam} X^{T,\Lam}_{-k_3,k_2}(k_1+k_2) &=&
 - \frac{1}{4}  
 \Pi^{PP}_{++-+}(k_1,k_2,k_3,k_1+k_2-k_3) \, .
 \hskip 1cm
\end{eqnarray}
Eqs.~(\ref{xdirect}) and (\ref{xcrossed}) are equivalent, since
\begin{equation}
 \Pi^{PH,d}_{s_1s_2s_3s_4}(k_1,k_2,k_3,k_4) =
 \Pi^{PH,cr}_{s_2s_1s_3s_4}(k_2,k_1,k_3,k_4) \; .
\end{equation}
Solving for $X^{S,\Lam}$ and $X^{T,\Lam}$, and relabeling
variables, one obtains
\begin{eqnarray}
 \frac{d}{d\Lam} X^{S,\Lam}_{k_1,k_2}(q) &=&
 \Pi^{PH,d}_{++-+}(k_1,k_2,k_2-q,k_1+q) +
 \frac{1}{4} \Pi^{PP}_{++-+}(q-k_2,k_2,-k_1,k_1+q) \; , 
 \nonumber \\ \\
  \frac{d}{d\Lam} X^{T,\Lam}_{k_1,k_2}(q) &=&
 - \frac{1}{4} \Pi^{PP}_{++-+}(q-k_2,k_2,-k_1,k_1+q) \; .
\end{eqnarray}

For the anomalous (4+0)-interactions we expect a singular
dependence on the momentum of each Cooper pair contained
in $\Gam^{(4+0)}[\psi,\psib]$, Eq.~(\ref{gamma40}),
which leads us to the ansatz
\begin{eqnarray}
 W^{S,\Lam}(k_1,k_2,k_3,k_4) &=& 
 W^{S,\Lam}_{k_1,-k_3}(k_1+k_2) \; , \nonumber \\
 W^{T,\Lam}(k_1,k_2,k_3,k_4) &=& 
 W^{T,\Lam}_{k_1,-k_3}(k_1+k_2) \; ,
\end{eqnarray}
where $k_4 = - k_1 - k_2 - k_3$.
The corresponding Nambu component becomes
\begin{eqnarray}
 \Gam^{(4),\Lam}_{++--}(k_1,k_2,k_3,k_4) &=&
 W^{\Lam}(k_1,k_2,k_3,k_4) \nonumber \\
 &=& W^{S,\Lam}_{k_1,k_3}(k_3-k_2) +
  W^{T,\Lam}_{k_1,k_3}(k_3-k_2) \nonumber \\
 &-& W^{S,\Lam}_{k_1,k_4}(k_1-k_3) -
  W^{T,\Lam}_{k_1,k_4}(k_1-k_3) +
  2W^{T,\Lam}_{k_1,k_3}(k_1+k_2) \, , \hskip 1cm
\end{eqnarray}
where $k_4 = k_1 + k_2 - k_3$.
Comparing once again the momentum dependences with those 
of the bubbles, we assign
\begin{eqnarray}
 \frac{d}{d\Lam} \left[ W^{S,\Lam}_{k_1,k_3}(k_3-k_2) +
  W^{T,\Lam}_{k_1,k_3}(k_3-k_2) \right] &=&
 \Pi^{PH,d}_{++--}(k_1,k_2,k_3,k_1 \!+\! k_2 \!-\! k_3) \; ,
 \label{wdirect} \\
 \frac{d}{d\Lam} \left[ W^{S,\Lam}_{k_1,k_4}(k_1-k_3) +
  X^{T,\Lam}_{k_1,k_4}(k_1-k_3) \right] &=&
 \Pi^{PH,cr}_{++--}(k_1,k_2,k_3,k_1 \!+\! k_2 \!-\! k_3) \; ,
 \label{wcrossed} \\
 \frac{d}{d\Lam} W^{T,\Lam}_{k_1,k_3}(k_1+k_2) &=&
 - \frac{1}{4}  
 \Pi^{PP}_{++--}(k_1,k_2,k_3,k_1 \!+\! k_2 \!-\! k_3) \, ,
 \hskip 1cm
\end{eqnarray}
where Eqs.~(\ref{wdirect}) and (\ref{wcrossed}) are 
equivalent due to exchange symmetry.
Solving for $W^{S,\Lam}$ and $W^{T,\Lam}$, and relabeling
variables, one obtains
\begin{eqnarray}
 \frac{d}{d\Lam} W^{S,\Lam}_{k_1,k_2}(q) &=&
 \Pi^{PH,d}_{++--}(k_1,k_2-q,k_2,k_1-q) +
 \frac{1}{4} \Pi^{PP}_{++--}(k_1,q-k_1,k_2,q-k_2) \; , 
 \nonumber \\ \\
  \frac{d}{d\Lam} W^{T,\Lam}_{k_1,k_2}(q) &=&
 - \frac{1}{4} \Pi^{PP}_{++--}(k_1,q-k_1,k_2,q-k_2) \; .
\end{eqnarray}

So far no approximation has been made in rewriting the
flow equations. Consequently, they capture the exact flow 
of the reduced (mean-field) model discussed in {\S}~4.
There, however, redundancies in the Nambu vertex for the
reduced model could be exploited in order to construct the
solution entirely in the momentum channel $k_1 = k_4$,
$k_2 = k_3$, where the flow is determined exclusively 
by the direct Nambu particle-hole diagram. 

The channel decomposition of the vertex and the flow 
equations provides a very useful starting point for an 
efficient approximate parametrization of the momentum and 
energy dependences being generated in the various channels 
in the course of the flow for models with generic 
interactions, as shown for the normal (not symmetry-broken) 
state of an interacting Fermi system by Husemann and 
Salmhofer.\cite{Husemann09}

\section{Conclusions}

We have addressed the problem of finding an efficient
parametrization for the effective two-particle vertex in
a spin-singlet superconductor, with the perspective to 
solve functional renormalization group flow equations for
interacting Fermi systems with a superconducting ground
state. 
We have constructed a manifestly spin-rotation invariant
form of the vertex, which reduces the number of independent
Nambu components to only three functions ($V$, $W$, and $X$).
By studying the exact flow of the vertex for a reduced
(mean-field) model exhibiting superconductivity and also
forward scattering, we have identified the singularities of
the vertex associated with the superconducting instability.
We have then expressed the vertex as a sum of various 
interaction channels where potential singularities are isolated 
in only one momentum and frequency variable in each channel,
and derived the corresponding channel-decomposed flow 
equations.

Our work paves the way for a controlled solution of the rather 
complex flow equation governing the flow of an interacting
Fermi system into a superconducting phase. Since singular
dependences generated by fermion loops have been isolated in 
only one momentum and energy variable per channel, one can 
parametrize these singularities by a relatively simple ansatz 
with a tractable number of parameters.

\section*{Acknowledgements}

We would like to thank J. Bauer, C. Honerkamp, C. Husemann 
and M. Salmhofer for valuable discussions, and S. Takei for
useful comments on the manuscript.
This work was supported by the German Research Foundation
through the research group FOR 723.

\end{document}